\documentclass[a4paper,10pt]{article}
\usepackage{subfigure}
\usepackage{amsfonts}
\usepackage[dvips]{color,graphicx}
\usepackage{graphicx,amssymb,amsmath,bm,latexsym}
\usepackage{stmaryrd}
\usepackage{bm}
\usepackage{CJK}
\usepackage{indentfirst}
\usepackage{amsmath}

\definecolor{red  }{rgb}{1,0,0}
\definecolor{blue }{rgb}{0,0,1}
\definecolor{green}{rgb}{0,1,0}

\usepackage{graphicx,amssymb,amsmath,bm,latexsym}

\usepackage{CJK}
\begin{document}
\begin{CJK*}{GBK}{song}
\vspace{-2.3em}
\title{Painlev\'e property, local and nonlocal symmetries and symmetry reductions for a (2+1)-dimensional integrable KdV equation}
\author{Xiao-bo Wang, Man Jia and S. Y. Lou\thanks{Corresponding author:lousenyue@nbu.edu.cn. Data Availability Statement: The data that support the findings of this study are available from the corresponding author upon reasonable request.}\\
\footnotesize \it School of Physical Science and Technology, Ningbo University, Ningbo, 315211, China}
\date{\today}
\maketitle

\vspace{-1.5em}
\begin{abstract}
The Painlev\'e property for a (2+1)-dimensional Korteweg-de Vries (KdV) extension, the combined KP3 (Kadomtsev- Petviashvili) and KP4 (cKP3-4) is proved by using Kruskal's simplification. The truncated Painlev\'e expansion is used to find the Schwartz form, the B\"acklund/Levi transformations and the residual nonlocal symmetry. The residual symmetry is localized to find its finite B\"acklund transformation. The local point symmetries of the model constitute a centerless Kac-Moody-Virasoro algebra. The local point symmetries are used to find the related group invariant reductions including a new Lax integrable model with a fourth order spectral problem. The finite transformation theorem or the Lie point symmetry group is obtained by using a direct method.
\end{abstract}
{\bf Keywords: \rm Painlev\'e property, residual symmetry, Schwartz form, B\"acklund transforms, D'Alembert waves, symmetry reductions, Kac-Moody-Virasoro algebra, (2+1)-dimensional KdV equation }
\\
{ \small PACS:05.45.Yv, 02.30.Ik, 47.20.Ky, 52.35.Mw, 52.35.Sb}

\section{Introduction \vspace{-0.3em}}
Recently, a novel (2+1)-dimensional Korteweg-de Vries (KdV) extension, the combined KP3 (Kadomtsev- Petviashvili) and KP4 (cKP3-4) equation
\begin{eqnarray}
&&u_{xt}=a[(6uu_x+u_{xxx})_x-3u_{yy}]+b\left(2vu_x+v_{xxx}+4uu_y\right)_x-v_{yy}, \label{kp34}\\
&&u_y=v_x. \nonumber
\end{eqnarray}
is proposed by one of the present authors (Lou) \cite{KP34}. KdV equation \cite{KdV} and its (2+1)-dimensional extensions such as the KP equation \cite{KP}, the Nizhnik-Novikov-Veselov (NNV)
equation\cite{NNV1,NNV2,NNV3}, the asymmetric NNV equation (ANNV) \cite{ANNV1,ANNV2,ANNV3} and the
Ito equation are fundamental nonlinear integrable models in mathematical physics \cite{Ito}.

The Lax integrability of the cKP3-4 equation is guaranteed by the existence of the Lax pair \cite{KP34} ($w_x=v_y$)
\begin{eqnarray}
\psi_{y}&=&\mbox{\rm i}(\psi_{xx}+u\psi),\ \mbox{\rm i}\equiv \sqrt{-1}, \label{psiy}\\
\psi_{t}&=&2\mbox{\rm i}b\psi_{xxxx}+4a\psi_{xxx}+4\mbox{\rm i}bu\psi_{xx}+2(3au+2\mbox{\rm i}bu_x
+bv)\psi_x\nonumber\\
&&-\mbox{\rm i}(3av+bw-2bu^2+3a\mbox{\rm i}u_x-2bu_{xx}+\mbox{\rm i}bv_x)\psi,\label{psit}
\end{eqnarray}
and the dual Lax pair
\begin{eqnarray}
\phi_{y}&=&-\mbox{\rm i}(\phi_{xx}+u\phi),\  \label{phiy}\\
\phi_{t}&=&-2\mbox{\rm i}b\phi_{xxxx}+4a\phi_{xxx}-4\mbox{\rm i}bu\phi_{xx}+2(3au-2\mbox{\rm i}bu_x
+bv)\phi_x\nonumber\\
&&+\mbox{\rm i}(3av+bw-2bu^2-3a\mbox{\rm i}u_x-2bu_{xx}-\mbox{\rm i}bv_x)\phi.\label{phit}
\end{eqnarray}

In Ref. \cite{KP34}, the multiple solitons of the model \eqref{kp34} are obtained by using Hirota's bilinear approach. Applying the velocity resonant mechanism \cite{JPCommun,Xu,Yan} to the multiple soliton solutions, the soliton molecules with arbitrary number of solitons are also found in \cite{KP34}. It is further discovered that the model permits the existence of the arbitrary D'Alembert type waves which implies that there are one special type of solitons and soliton molecules with arbitrary shapes but fixed model dependent velocity.

In this paper, we investigate other significant properties such as the Painlev\'e property (PP), Schwartz form, B\"aclund transformations, infinitely many local and nonlocal symmetries, Kac-Moody-Virasoro symmetry algebras, group invariant solutions and symmetry reductions for the cKP3-4 equation \eqref{kp34}. To study the PP of a nonlinear partial differential equation system, there are some equivalent ways such as the Weiss-Tabor-Carnevale (WTC) approach \cite{WTC}, Kruskal's simplification, Conte's invariant form \cite{Conte} and Lou's extended method \cite{LouE}. In the section 2 of this paper, the PP of \eqref{kp34} is tested by using the Kruskal's simplification. Using the truncated Panlev\'e expansion, one can find many interesting results for integrable systems including the B\"acklund/Levi transformation, Schwarz form, bilinearization and Lax pair. In Ref. \cite{GXN}, it is found that the nonlocal symmetries, the residual symmetries can also be directly obtained from the truncated Painlev\'e expansion. The residual symmetries can be used to find Dabourx transformations \cite{DT,DT1} and the interaction solutions between a soliton and another nonlinear wave such as a cnoidal wave and/or a Painlev\'e wave \cite{Int,Int1}. In the section 3, the nonlocal symmetry (the residual symmetry) is localized by introducing a prolonged system. Whence a nonlocal symmetry is localized, it is straightforward to find its finite transformation which is equivalent to the B\"acklund/Levi transformation. In section 4, it is found that similar to the usual KP equation, the general Lie point symmetries of the cKP3-4 equation possess also three arbitrary functions of the time $t$ and constitute a centerless Kac-Moody-Virasoro symmetry algebra. Using the general Lie point symmetries, two special types of symmetry reductions are found. The first type of (1+1)-dimensional reduction equation is Lax integrable with fourth order spectral problem. The second type of symmetry reduction equation is just the usual KdV equation. In section 5, we study the finite transformation theorem of the general Lie point symmetries via a simple direct method instead of the traditional complicated method by solving an initial value problem. The last section includes a short summary and some discussions.
\section{Painlev\'e property, B\"acklund transformation and Schwartz form of the cKP3-4 equation}
According to the standard WTC approach, if the model \eqref{kp34} is Painlev\'e integrable, all the possible solutions of the model can be written
as
\begin{eqnarray}
&&u=\sum_{j=0}^{\infty}u_{j}\phi^{j-\alpha},\ v=\sum_{j=0}^{\infty}v_{j}\phi^{j-\beta},\label{uv}
\end{eqnarray}
with four arbitrary functions among $u_j$ and $v_j$ in addition to the fifth arbitrary function, the arbitrary singular manifold $\phi$, where $\alpha$ and $\beta$
 should be positive integers. In other words, all the solutions of the model are single valued about the arbitrary movable singular manifold $\phi$.

To fix the constants $\alpha$ and $\beta$, one may use the standard leading order analysis. Substituting $u\sim u_0\phi^{-\alpha}$ and $ v\sim v_0\phi^{-\beta}$ into \eqref{kp34},
 and comparing the leading terms for
$\phi \sim 0$, we get the only possible branch with
\begin{eqnarray}
u_0=-2\phi_{x}^2,\ v_2=-2\phi_x\phi_y,\ \alpha=\beta=2.\label{uv0}
\end{eqnarray}
Substituting \eqref{uv} with \eqref{uv0} into \eqref{kp34} yields the recursion relation on the functions $\{u_j,\ v_j\}$
\begin{eqnarray}
\left(\begin{array}{cc}
   J_{11} & J_{12}\\
 (j-2)f_y & -(j-2)f_x
\end{array}\right)
\left(\begin{array}{c}
 u_j   \\ v_j
\end{array}\right)\equiv J\left(\begin{array}{c}
 u_j   \\ v_j
\end{array}\right)= \left(\begin{array}{c}
 F_1(u_0,\ u_1,\ \ldots,\ u_{j-1},\ v_0,\ v_1,\ \ldots,\ v_{j-1})   \\
  F_2(u_0,\ u_1,\ \ldots,\ u_{j-1},\ v_0,\ v_1,\ \ldots,\ v_{j-1})
\end{array}\right),\label{recur}
\end{eqnarray}
where $J_{11}=(j-5)(j-6)\left[4bf_x^3f_y-a(j+1)(j-4)f_x^4\right],\ J_{12}=-bj(j-3)(j-5)(j-6)f_x^4$, $F_1$ and $F_2$ are dependent only on $u_0,\ u_1,\ \ldots,\ u_{j-1},\ v_0,\ v_1,\ \ldots,\ v_{j-1}$ and the derivatives of $\phi$ with respect to $x,\ y$ and $t$. The determinant of the matrix $J$ reads
\begin{eqnarray}
\det{J}=f_x^3(j+1)(j-2)(j-4)(j-5)(j-6)(af_x+bf_y).\label{detJ}
\end{eqnarray}
From \eqref{recur} and \eqref{detJ}, we have
\begin{eqnarray}
\left(\begin{array}{c}
 u_j   \\ v_j
\end{array}\right)=J^{-1}\left(\begin{array}{c}
 F_1   \\ F_2
\end{array}\right)\label{uvj}
\end{eqnarray}
for $j\neq -1,\ 2,\ 4,\ 5$ and $6$. At the resonance points $j=-1,\ 2,\ 4,\ 5$ and $6$, five arbitrary functions, say, $\{\phi,\ u_2,\ u_4,\ u_5,\ u_6\}$, can be included in the formal solution \eqref{uv} if all the resonant conditions from \eqref{recur} are satisfied. To verify these resonant conditions, one can use the Kruskal's simplification for the singular manifold $\phi\sim 0$ replaced by $\phi\sim x+\psi(y,\ t)\sim 0$. Under the Krudkal's simplification, it is straightforward to find
\begin{eqnarray}
&&u_0=-2,\ v_0=-2\psi_y,\ u_1=v_1=0,\ u_3=-\frac12\psi_{yy},\nonumber \\
&&v_2=\frac{3a}{2b}\left(\psi_y^2-2u_2\right)+\frac12\psi_y^3-2u_2\psi_y+\frac1{2b}\psi_t,\ v_3=-\frac12\psi_{yy}\psi_y+u_{2y},\nonumber \\
&&v_4 = u_4\psi_y-\frac14\psi_{yyy},\ v_5 = u_5\psi_y+\frac13u_{4y},\ v_6 = u_6\psi_y+\frac14u_{5y}
\end{eqnarray}
while $\psi,\ u_2,\ u_4,\ u_5$ and $u_6$ are arbitrary functions of $y$ and $t$.
Thus the model \eqref{kp34} is not only Lax integrable but also Painlev\'e integrable.

Because the cKP3-4 equation \eqref{kp34} is Painlev\'e integrable, we can use the truncated Painelev\'e expansion,
\begin{eqnarray}
&&u=\frac{u_{0}}{\phi^{2}}+\frac{u_{1}}{\phi}+u_2,\ v=\frac{v_{0}}{\phi^{2}}+\frac{v_{1}}{\phi}+v_2,\label{uv20}
\end{eqnarray}
to find other interesting properties of the cKP3-4 equation \eqref{kp34}.

It is known that using the relations \eqref{uv20} with $u_2=v_2=0$, the cKP3-4 equation can be bilinearized \cite{KP34} to
\begin{eqnarray}
[D_xD_{\tau} +a(3 D_y^2-D_x^4)]f\cdot f=0, \label{BKP}
\end{eqnarray}
and
\begin{equation}
[a(2bD_x^3D_y-3D_xD_t+3D_xD_{\tau})+bD_yD_{\tau}]f\cdot f=0, \label{JM}
\end{equation}
with help of the auxiliary variable $\tau$, where the Hirota's bilinear operators $D_z,\ z=x,\ y,\ t,\ \tau$ are defined by
\begin{eqnarray}
D_z^n f\cdot g= \left.(\partial_z-\partial_{z'})^n f(z)g(z')\right|_{z'=z}. \label{Dz}
\end{eqnarray}

After introducing M\"obious transformation ($\phi \rightarrow \frac{c_0+c_1\phi}{b_0+b_1\phi}$ with $c_0b_1\neq c_1b_0$) invariants,
\begin{eqnarray}
g=\frac{\phi_t}{\phi_x},\ h=\frac{\phi_y}{\phi_x},\ S=\frac{\phi_{xxx}}{\phi_x}-\frac32\frac{\phi_{xx}^2}{\phi_x^2},\ w=a(S_x-3h_y-3hh_x)-g_x,\label{sg}
\end{eqnarray}
and substituting \eqref{uv20} with $u_2v_2\neq 0$ into \eqref{kp34}, one can directly obtain the auto and/or non-auto B\"acklund transformation (BT) theorem and the residual symmetry theorem:\\
\bf Theorem 1. \it B\"acklund transformation theorem. \rm
If $\phi$ is a solution of the Schwartz cKP3-4 equation
\begin{eqnarray}
&&h_{xx}^2(h_{xx}^{-1}w_x)_x=b\big[S_x(3h_xh_{xxx}-5h_{xx}^2)+S_{2x}(hh_{xxx}-4h_xh_{xx})
-hh_{xx}S_{xxx}-h_{xx}h_{xxxxx}+h_{xxx}h_{xxxx}\nonumber\\
&&\quad -(3h_xh_y+hh_{xy}+h_{yy})h_{xxx}+5h_yh_{xx}^2
+h_{xx}(4h_xh_{xy}+hh_{xxy}+h_{yyx})\big],\label{SKP34}
\end{eqnarray}
then both
\begin{eqnarray}\left\{\begin{array}{l}
u_a=\frac{h^2}4+\frac{aw_x}{4bh_{xx}}-S-\frac{\phi_{xx}^2}{\phi_x^2}
-\frac{3h_xS_s+hS_{xx}+h_{xxxx}-hh_{xy}-h_{yy}+3h_xh_y}{4h_{xx}},\\
v_a=\frac{g+bh^3+3ah^2}{2b}-\frac2b(a+bh)S-\frac{3a+2bh}b u_a-h_{xx}
-\frac{3(a+bh)\phi_{xx}^2}{\phi_x^2}-\frac{h_x\phi_{xx}}{\phi_x}
\end{array}\right.\label{uva}
\end{eqnarray}
and
\begin{eqnarray}\left\{\begin{array}{l}
u_b=u_a-\frac{2\phi_x^2}{\phi^2}+\frac{2\phi_{xx}}{\phi},\\
v_b=v_a-\frac{2\phi_x\phi_y}{\phi^2}+\frac{2\phi_{xy}}{\phi},
\end{array}\right.\label{uvb}
\end{eqnarray}
are solutions of the cKP3-4 equation \eqref{kp34}.\\
\bf Theorem 2. \it Residual symmetry theorem. \rm If $\phi$ is a solution of the Schwartz cKP3-4 equation \eqref{SKP34}, and the fields $\{u,\ v\}=\{u_a,\ v_a\}$ are related to the singular manifold $\phi$ by \eqref{uva}, then
\begin{equation}
\{\sigma^u,\ \sigma^v\}=\{2\phi_{xx},\ 2\phi_{xy}\} \label{uvs}
\end{equation}
is a nonlocal symmetry (residual symmetry) of the cKP3-4 equation \eqref{kp34}.
In other words, \eqref{uvs} solve the symmetry equations, the linearized equations of \eqref{kp34}
\begin{eqnarray}
&&\sigma^u_{xt}=a[(6\sigma^uu_x+6u\sigma^u_x+\sigma^u_{xxx})_x-3\sigma^u_{yy}]
+b\left(2v\sigma^u_x+2\sigma^vu_x+\sigma^v_{xxx}+4\sigma^uu_y+4u\sigma^u_y\right)_x-\sigma^v_{yy}, \label{kp34sym}\\
&&\sigma^u_y=\sigma^v_x. \nonumber
\end{eqnarray}
From \eqref{SKP34}, one can find that when $b=0$, the Schwartz cKP3-4 is reduced back to the usual Schwartz KP equation
$$w=0.$$
 The BT \eqref{uva} is a non-auto BT because it changes a solution of the Schwartz cKP3-4 equation \eqref{SKP34} to that of the usual cKP3-4 equation \eqref{kp34}. The BT \eqref{uvb} may be considered as a non-auto BT if $u_a$ and $v_a$ are replaced by \eqref{uva}. The BT \eqref{uvb} may also be considered as an auto-BT which changes one solution $\{u_a,\ v_a\}$ to another $\{u_b,\ v_b\}$ for the same equation \eqref{kp34}.

From the auto-BT \eqref{uvb} and the trivial seed solution $\{u_a=0,\ v_a=0\}$, one can obtain some interesting exact solutions. Substituting $\{u_a=0,\ v_a=0\}$ into \eqref{uva}, we have
\begin{eqnarray}
&&\frac{h^2}4+\frac{aw_x}{4bh_{xx}}-S-\frac{\phi_{xx}^2}{\phi_x^2}
-\frac{3h_xS_s+hS_{xx}+h_{xxxx}-hh_{xy}-h_{yy}+3h_xh_y}{4h_{xx}}=0,\label{ua0}\\
&&\frac{g+bh^3+3ah^2}{2b}-\frac2b(a+bh)S-h_{xx}
-\frac{3(a+bh)\phi_{xx}^2}{\phi_x^2}-\frac{h_x\phi_{xx}}{\phi_x}.\label{va0}
\end{eqnarray}
After solving the over determined system \eqref{SKP34}, \eqref{ua0} and \eqref{va0}, one can find various exact solutions from the BT \eqref{uvb} with $\{u_a=0,\ v_a=0\}$. Here, we discuss only for the travelling wave solutions of the system \eqref{SKP34}, \eqref{ua0} and \eqref{va0}. For the travelling wave, $\phi=\Phi(kx+py+\omega t)$, the Schwartz equation \eqref{SKP34} becomes an identity while \eqref{ua0} and \eqref{va0} becomes
\begin{eqnarray}
&&(ak+bp)[\Phi_{\xi\xi\xi\xi\xi}\Phi_{\xi}^3
-(5\Phi_{\xi\xi\xi\xi}\Phi_{\xi\xi}+4\Phi_{\xi\xi\xi}^2)\Phi_{\xi}^2
+17\Phi_{\xi}\Phi_{\xi\xi}^2\Phi_{\xi\xi\xi}-9\Phi_{\xi\xi}^4]=0,\label{TWua0}\\
&&(3akp^2+bp^3+k^2\omega)\Phi_{\xi}^2-k^4(ak+bp)(4\Phi_{\xi}\Phi_{\xi\xi\xi}-3\Phi_{\xi\xi}^2)=0.
\label{TWva0}
\end{eqnarray}
Here we list three special solution examples of the cKP3-4 equation \eqref{kp34} related to \eqref{TWua0} and \eqref{TWva0}. \\
\bf Example 1. \it D'Alembert type arbitrary travelling waves moving in one direction with a fixed model dependent velocity. \rm
$$\Phi=\Phi(\xi),\ \xi=b^2x -2a^3t-aby, \ p = -ak/b,\ \omega = -2ka^3/b^2,$$
\begin{eqnarray}
&&u=-bv/a= 2b^4[\ln(\Phi)]_{\xi\xi},
\label{DW}
\end{eqnarray}
where $\Phi$ is an arbitrary function of $\xi=b^2x -2a^3t-aby$.

Because of the arbitrariness of $\Phi$, the localized excitations with special fixed model dependent velocity $\{-2a^3/b^2,\ -2a^2/b\}$ possess rich structures including kink shapes, plateau shapes, molecule forms, few cycle forms, periodic solitons, etc. in addition to the usual $\mbox{\rm sech}^2$ form \cite{KP34}.\\
\bf Example 2. \it Rational wave. \rm
$$\Phi=k x+p y-p^2 k^{-2} (3 a k+b p) t+\xi_0,$$
\begin{eqnarray}
&&u=kv/p= -\frac{2 k^6}{(3 a k p^2 t+b p^3 t-k^3 x-k^2 p y-\xi_0 k^2)^2}
\label{RW}
\end{eqnarray}
with arbitrary constants $k,\ p$ and $\xi_0$.\\
\bf Example 3. \it Soliton solution. \rm
$$\Phi=1+\exp(\xi),\ \xi=kx+py-\frac{1}{k^2}(-ak^5 -bk^4p+3akp^2+bp^3)t+\xi_0$$
\begin{eqnarray}
&&u=kv/p= \frac{k^2}2\mbox{\rm sech}^2\left(\frac{\xi}2\right),
\label{SW}
\end{eqnarray}
with arbitrary constants $k,\ p$ and $\xi_0$.

Different from the D'Alembert wave \eqref{DW}, the soliton solution \eqref{SW} possesses arbitrary velocity $\{-p/k,$ $ -ak^2 -bkp+3ak^{-2}p^2+bp^3k^{-3}\}$ but fixed $\mbox{\rm sech}^2$ shape.

\section{Localization of nonlocal symmetry \eqref{uvs}}
Similar to the usual KP equation \cite{Int1} and the supersymmetric KdV equation \cite{superKdV}, the nonlocal symmetry (residual symmetry) \eqref{uvs} can be localized by introducing auxiliary variables
\begin{equation}
\phi_1=\phi_x,\ \phi_2=\phi_y,\ \phi_3=\phi_{1x},\ \phi_{4}=\phi_{2x}. \label{phi123}
\end{equation}
It is straightforward to verify that the nonlocal symmetry of the cKP3-4 equation \eqref{kp34} becomes a local one for the prolonged system \eqref{kp34}, \eqref{SKP34}, \eqref{uva} with $\{u_a=u,\ v_a=v\}$ and \eqref{phi123}. The vector form of the localized symmetry of the prolonged system can be written as
\begin{equation}
V=2\phi_3\partial_u+2\phi_4\partial_v-\phi^2\partial_{\phi}-2\phi \phi_1 \partial_{\phi_1}
-2\phi \phi_2\partial_{\phi_2}-2(\phi_1^2+\phi\phi_3)\partial_{\phi_3} -2(\phi_1\phi_2+\phi\phi_4)\partial_{\phi_4}. \label{V}
\end{equation}
According to the closed prolongation structure \eqref{V}, one can readily obtain the finite transformation (auto B\"acklund transformation) theorem by solving the initial value problem
\begin{eqnarray}
&&\frac{\mbox{\rm d} u(\epsilon)}{\mbox{\rm d} \epsilon}=2\phi_3(\epsilon),\  \frac{\mbox{\rm d} v(\epsilon)}{\mbox{\rm d} \epsilon}=2\phi_4(\epsilon),\
\frac{\mbox{\rm d} \phi(\epsilon)}{\mbox{\rm d} \epsilon}=-\phi(\epsilon)^2,\
\frac{\mbox{\rm d} \phi_1(\epsilon)}{\mbox{\rm d} \epsilon}=-2\phi(\epsilon) \phi_1(\epsilon),\
\frac{\mbox{\rm d} \phi_2(\epsilon)}{\mbox{\rm d} \epsilon}=-2\phi(\epsilon) \phi_2(\epsilon),\
\nonumber\\
&&\frac{\mbox{\rm d} \phi_3(\epsilon)}{\mbox{\rm d} \epsilon}=-2[\phi_1(\epsilon)^2+\phi(\epsilon)\phi_3(\epsilon)],\
\frac{\mbox{\rm d} \phi_4(\epsilon)}{\mbox{\rm d} \epsilon}=-2[\phi_1(\epsilon)\phi_2(\epsilon)+\phi(\epsilon)\phi_4(\epsilon)], \label{IVP}\\
&&\left.\{u(\epsilon),\ v(\epsilon),\ \phi(\epsilon),\ \phi_1(\epsilon),\ \phi_2(\epsilon),\ \phi_3(\epsilon),\ \phi_4(\epsilon)\}\right|_{\epsilon=0}=\{u,\ v,\ \phi,\ \phi_1,\ \phi_2,\ \phi_3,\ \phi_4\}.\nonumber
\end{eqnarray}
\bf Theorem 3. \it Auto B\"acklund transformation theorem. \rm If $\{u,\ v,\ \phi,\ \phi_1,\ \phi_2,\ \phi_3,\ \phi_4\}$ is a solution of the prolonged system \eqref{kp34}, \eqref{SKP34}, \eqref{uva} with $\{u_a=u,\ v_a=v\}$ and \eqref{phi123}, so is $\{u(\epsilon),\ v(\epsilon),\ \phi(\epsilon),\ \phi_1(\epsilon),\ \phi_2(\epsilon),\ \phi_3(\epsilon),\ $ $ \phi_4(\epsilon)\}$ with
\begin{eqnarray}
&&\phi(\epsilon)=\frac{\phi}{1+\epsilon \phi},\ \phi_1(\epsilon)=\frac{\phi_1}{(1+\epsilon \phi)^2},\ \phi_2(\epsilon)=\frac{\phi_2}{(1+\epsilon \phi)^2},\nonumber\\
&& \phi_3(\epsilon)=\frac{\phi_3}{(1+\epsilon \phi)^2}-\frac{2\epsilon\phi_1^2}{(1+\epsilon \phi)^3},\
\phi_4(\epsilon)=\frac{\phi_4}{(1+\epsilon \phi)^2}-\frac{2\epsilon\phi_1\phi_2}{(1+\epsilon \phi)^3},\label{BTepsilon}\\
&&
u(\epsilon)=u+\frac{2\epsilon\phi_3}{1+\epsilon \phi}-\frac{2\epsilon^2\phi_1^2}{(1+\epsilon \phi)^2},\
v(\epsilon)=v+\frac{2\epsilon\phi_4}{1+\epsilon \phi}-\frac{2\epsilon^2\phi_1\phi_2}{(1+\epsilon \phi)^2}. \label{AutoBT}
\end{eqnarray}
Comparing the theorem 2 and the theorem 3, one can find that for the cKP3-4 equation \eqref{kp34}, the transformation \eqref{AutoBT} is equivalent to \eqref{uvb} by using the transformation $1+\epsilon \phi\rightarrow \phi$.

\section{Symmetry reductions of the cKP3-4 equation}
Using the standard Lie point symmetry method or the formal series symmetry approach \cite{PRL_Lou,JPA_Lou} to the cKP3-4 equation, it is straightforward to find the general Lie point symmetry solutions of \eqref{kp34sym} are generated by the following three generators,
\begin{eqnarray}
\left(\begin{array}{c}\sigma^u \\ \sigma^v \end{array}\right)=K_0(\alpha)=\left(\begin{array}{c} \alpha u_x \\ \alpha v_x+\frac1{2b}\alpha_t \end{array}\right),
\label{alpha}
\end{eqnarray}
\begin{eqnarray}
\left(\begin{array}{c}\sigma^u \\ \sigma^v \end{array}\right)=K_1(\beta)=\left(\begin{array}{c} \beta u_y+\frac1{4b}\beta_t \\  \beta v_y-\frac{3a}{4b^2}\beta_t\end{array}\right),
\label{beta}
\end{eqnarray}
and
\begin{eqnarray}
K_2(\theta)=\left(\begin{array}{c} \theta u_t+\frac{\theta_t}{4b}(ay+bx)u_x+\frac{1}2\theta_t (yu)_y+\frac{3a^2}{8b^2}\theta_t+\frac{y}{8b}\theta_{tt} \\
\theta v_t+\frac{\theta_t}{4b}(ay+bx)v_x -\frac{1}{2}\theta_t(yv)_y +\frac{au+bv}{4b}\theta_{t}+\frac{bx-2ay}{8b^2}\theta_{tt}-\frac{9a^3}{8b^3}\theta_t
\end{array}
\right),\label{theta}
\end{eqnarray}
where $\alpha,\ \beta$ and $\theta$ are arbitrary functions of $t$.

The symmetries $K_0(\alpha),\ K_1(\beta)$ and $K_0(\theta)$ constitute a special Kac-Moody-Virasoro algebra with the nonzero commutators
\begin{eqnarray}
[K_2(\theta),\ K_0(\alpha)]=K_0(\theta\alpha_t),\ [K_2(\theta),\ K_1(\beta)]=K_1(\theta\beta_t),\
[K_2(\theta_1),\ K_2(\theta_2)]=K_2(\theta_1\theta_{2t}-\theta_{1t}\theta_2),
\label{K20}
\end{eqnarray}
where the commutator $[F,\ G]$ with $F=\big(F_1(u,\ v), F_2(u,\ v)\big)^{T}$ and $G=\big(G_1(u,\ v), G_2(u,\ v)\big)^T$, where the superscript $T$ means the transposition of matrix, is defined by
$$[F,\ G]\equiv \left(\begin{array}{cc}
F'_{1u} & F'_{1v}\\
F'_{2u} & F'_{2v}
\end{array}\right)G-\left(\begin{array}{cc}
G'_{1u} & G'_{1v}\\
G'_{2u} & G'_{2v}
\end{array}\right)F,
$$
and $F'_{1u},\ F'_{1v},\ F'_{2u},\ F'_{2v},\ G'_{1u},\ G'_{1v},\ G'_{2u}$ and $ G'_{2v}$ are partial linearized operators, say,
$$F'_{1u}G_1\equiv \left. \frac{\mbox{\rm d}}{\mbox{\rm d}\epsilon}F_1(u+\epsilon G_1,\ v)\right|_{\epsilon=0}.$$

From \eqref{K20}, we know that $K_0$ and $K_1$ constitute the usual Kac-Moody algebra and $K_2$ constitutes the Virasoro algebra if we fix the arbitrary functions $\alpha,\ \beta$ and $\theta$ as special exponential functions $\exp (mt)$ or polynomial functions $t^m$ for $m=0,\ \pm1,\ \pm2,\ \ldots.$

Applying the Lie point symmetries $K_0(\alpha),\ K_1(\beta)$ and $K_0(\theta)$ to the cKP3-4 equation \eqref{kp34}, we can get two nontrivial symmetry reductions. \\
\bf Reduction 1: $\theta\neq0$. \rm For $\theta\neq0$, we rewrite the arbitrary functions in the form
\begin{equation}
\theta\equiv \rho^4\neq0,\ \alpha\equiv \rho^5\alpha_{1t},\ \beta\equiv \rho^6\beta_{1t},\ \theta_{1t}\equiv \rho \beta_{1t}. \label{case1}
\end{equation}
Under the new definitions \eqref{case1}, the group invariant condition
$$K_0(\alpha)+K_1(\beta)+K_2(\theta)=0$$
yields the first type of group invariant solutions in the form
\begin{eqnarray}
&&u=\frac{U(\xi,\ \eta)}{\rho^{2}} -\frac{\rho_ty}{2b\rho}-\frac{3a^2}{4b^2}-\frac{\beta_{1t}\rho}{4b},\label{u1}\\
&&v=\frac{V(\xi,\ \eta)}{\rho^3}-\frac{aU(\xi,\ \eta)}{b\rho^2}+\frac{(2ay-bx)\rho_t}{2b^2\rho}+\frac{7a^3}{4b^3}-\frac{\rho\alpha_{1t}}{2b}
+\frac{3a\rho\beta_{1t}}{4b^2},\label{v1}
\end{eqnarray}
where $U(\xi, \ \eta)\equiv U$ and $V(\xi, \ \eta)\equiv V$ are group invariant functions of the group invariant variables $\xi$ and $\eta$ with
\begin{eqnarray}
\xi=\frac{x}{\rho}-\frac{ay}{b\rho}-\alpha_1+\frac{a\theta_1}b,\ \eta=\frac{y}{\rho^2}-\beta_1. \label{xieta}
\end{eqnarray}
Substituting \eqref{u1} and \eqref{v1} into \eqref{kp34}, we can find the group invariant reduction equations for the group invariant functions $U$ and $V$,
\begin{eqnarray}
&&U_{\eta}=V_{\xi},\nonumber\\
&&V_{\eta\eta}=(V_{\xi\xi\xi}+4UV_{\xi}+2VU_{\xi})_{\xi}. \label{RED1}
\end{eqnarray}
It is interesting that the reduction system \eqref{RED1} is Lax integrable with the fourth order spectral problem
\begin{eqnarray}
\lambda \Psi&=&2\Psi_{\xi\xi\xi\xi}+4U\Psi_{\xi\xi}+2(2U_{\xi}
-\mbox{\rm i}V)\Psi_{\xi}-\left(\int V_{\eta} \mbox{\rm d}\xi -2U^2-2U_{\xi\xi}+\mbox{\rm i}V_{\xi}\right)\Psi,\label{Psit}\\
\Psi_{\eta}&=&\mbox{\rm i}(\Psi_{\xi\xi}+U\Psi). \label{Psiy}
\end{eqnarray}

\bf Reduction 2: $\beta\neq0.$ \rm For  $\beta\neq0$ case, the group invariant condition
$$K_0(\alpha)+K_1(\beta)=0$$
and $v_x=u_y$ yield the usual KdV reduction
\begin{eqnarray}
(U_{T}+U_{XXX}+6UU_{X})_{X}=0,\ X=\frac1{\sqrt{\beta}}\left(\frac{\alpha}{\beta}y-x\right),\ T=\int\beta^{-5/2}(a\beta-b\alpha) \mbox{\rm d}t \label{RED2}
\end{eqnarray}
with
\begin{eqnarray}
&&u=\frac{U(X,T)}{\beta}-\frac{\beta_ty}{4b\beta}
+\frac{\alpha^2(3a\beta-b\alpha)}{6\beta^2(a\beta-b\alpha)},\label{u2}\\
&&v=\frac{-\alpha U(X,T)}{\beta^2}-\frac{\beta_tx}{4b\beta}
+\left[\frac{(3a\beta+b\alpha)\beta_t}{4b^2\beta^2}-\frac{\alpha_t}{2b\beta}\right]y
-\frac{\alpha^3(3a\beta-b\alpha)}{6\beta^3(a\beta-b\alpha)}.\label{v2}
\end{eqnarray}

\section{Finite transformation theorem of $K_0(\alpha)+K_1(\beta)+K_2(\theta)$ via direct method}
The finit transformation of $K_0(\alpha)+K_1(\beta)+K_2(\theta)$ may be obtained by solving the initial value problem ($\{x',\ y',\ t',\ u',\ v'\}=\{x'(\epsilon),\ y'(\epsilon),\ t'(\epsilon),\ u'(\epsilon),\ v'(\epsilon)\},\ \{\alpha', \ \beta',\ \theta'\}= \{\alpha(t'), \ \beta(t'),\ \theta(t')\}$),
\begin{eqnarray}
&&\frac{\mbox{\rm d}t'}{\mbox{\rm d}\epsilon}=\theta',\ \frac{\mbox{\rm d}y'}{\mbox{\rm d}\epsilon}=\left(\beta'+\frac12 \theta'_{t'}y'\right),\ \frac{\mbox{\rm d}x'}{\mbox{\rm d}\epsilon}=\left[\alpha'
+\frac1{4b}\theta'_{t'}(ay'+bx')\right],\label{x'}\\
&&\frac{\mbox{\rm d}u'}{\mbox{\rm d}\epsilon}=-\frac1{4b}\beta'_{t'}-\frac12u'\theta'_{t'}
-\frac1{8b}\theta'_{t't'}y'-\frac{3 a^2}{8 b^2}\theta'_{t'}, \\
&&\frac{\mbox{\rm d}v'}{\mbox{\rm d}\epsilon}=-\frac1{2b}\alpha'_{t'}+\frac{3a}{4b^2}\beta'_{t'}
-\frac{a}{4b}\theta'_{t'}u'-\frac34\theta'_{t'}v'+\frac{9 a^3}{8 b^3}\theta'_{t'}
+\frac{2ay'-bx'}{8b^2}\theta'_{t't'},\label{v'}\\
&&\ t'(0)=t,\ y'(0)=y, \ x'(0)=x,\ u'(0)=u,\  v'(0)=v. \nonumber
\end{eqnarray}
However, the exact solution of the initial value problem \eqref{x'}--\eqref{v'} is very complicated and quite awkward even for the pure KP ($a=0$) case \cite{Wint}. An alternative simple method is to find symmetry group via a direct method \cite{JMP03,JMP04,Ma1,Ma2} by using a priori ansatz
\begin{equation}
u=A+BU(x',\ y',\ t'),\ v=A_1+B_1U(x',\ y',\ t')+B_2V(x',\ y',\ t'), \label{uv'}
\end{equation}
to determine the functions $\{A,\ B,\ A_1,\ B_1,\ B_2, \ x',\ y', \ t'\}=\{A,\ B,\ A_1,\ B_1,\ B_2, \ x',\ y', \ t'\}(x,\ y, \ t)$ such that both $\{u,\ v\}$ and $\{U,\ V\}$ are solutions of the cKP3-4 equation \eqref{kp34}.

Substituting the ansatz \eqref{uv'} into \eqref{kp34} and requiring $U$ and $V$ satisfying the same cKP3-4 equation \eqref{kp34} but with different variables $\{x',\ y',\ t'\}$, one can readily determine all the undetermined functions $\{A,\ B,\ A_1,\ B_1,\ B_2,$ $ \ x',\ y', \ t'\}$. The final result can be summarized to the following finite transformation theorem. \\
\bf Theorem 4. \it Finite transformation theorem. \rm If $\{U,\ V\}=\{U(x,\ y,\ t),\ V(x,\ y,\ t)\}$ is a solution of \eqref{kp34}, so is $\{u,\ v\}$ with
\begin{eqnarray}
&&u=\sqrt{\tau_t}U(x',\ y',\ t')+\frac{\tau_{tt}y}{8b\tau_t} +\frac{y_{0t}}{4b\sqrt{\tau_t}}+\frac{3a^2}{4b^2}\big(\sqrt{\tau_t}-1\big),\label{U'}\\
&&v=\frac{a\sqrt{\tau_t}}{b}(\sqrt[4]{\tau_t}-1)U(x',\ y',\ t')+\sqrt[4]{\tau_t^3}V(x',\ y',\ t')+\frac{\tau_{tt}(bx-2ay)}{8b^2\tau_t}\nonumber\\
&&\qquad
+\frac{a^3}{4b^3}\left(7-3\sqrt{\tau_t}-4\sqrt[4]{\tau_t^3}\right)
+\frac{bx_{0t}-ay_{0t}}{2b^2\sqrt[4]{\tau_t}} -\frac{ay_{0t}}{4b^2\sqrt{\tau_t}}, \label{V'}\\
&&x'=\sqrt[4]{\tau_t}x-\frac{a}b(\sqrt[4]{\tau_t}-\sqrt{\tau_t})y+x_{0},\
y'=\sqrt{\tau_t}y+y_0,\ t'=\tau, \label{xyt'}
\end{eqnarray}
where $x_0=x_0(t),\ y_0=y_0(t)$ and $\tau=\tau(t)$ are three arbitrary functions of $t$.

To verify the correctness one can directly substitute \eqref{U'}--\eqref{xyt'} into \eqref{kp34}. In fact, one can take the arbitrary functions $x_0,\ y_0$ and $\tau$ in the forms
\begin{equation}
\tau=t+\epsilon \theta,\ x_0=\epsilon \alpha,\ y_0=\epsilon \beta, \label{xy0}
\end{equation}
where $\theta,\ \alpha$ and $\beta$ are arbitrary functions of $t$. Substituting \eqref{xy0} into \eqref{U'} and \eqref{V'} yields
\begin{equation}
\left(\begin{array}{c}
u \\ v
\end{array}\right)=\left(\begin{array}{c}
U \\ V
\end{array}\right)+\epsilon [K_2(\theta)+K_0(\alpha)+K_1(\beta)]+O(\epsilon^2), \label{K0K2}
\end{equation}
which means theorem 4 is just the finite transformation theorem of the symmetry $K_2(\theta)+K_0(\alpha)+K_1(\beta)$.

Applying theorem 4 to the D'Alembert wave \eqref{DW}, we get a new solution
\begin{eqnarray}
u&=&2b^4\sqrt{\tau_t}\left[\ln(\Phi)\right]_{\zeta\zeta}+\frac{\tau_{tt}y}{8b\tau_t}
+\frac{y_{0t}}{4b\sqrt{\tau_t}}+\frac{3a^2}{4b^2}\big(\sqrt{\tau_t}-1\big),\\
\zeta &=&b\sqrt[4]{\tau_t}\left(bx-ay
\right)+\zeta_0
\end{eqnarray}
with $\tau,\ y_0$ and $\zeta_0$ being arbitrary functions of $t$ and $\Phi$ being an arbitrary function of $\zeta$.

\section{Conclusions and discussions\vspace{-0.3em}}

In summary, the cKP3-4 equation \eqref{kp34} is a significant (2+1)-dimensional KdV extension with various interesting integrable properties. In this paper, the Painlev\'e property, auto- and nonauto- B\"acklund transformations, local and nonlocal symmetries, Kac-Moody-Virasoro symmetry algebra, finite transformations related to the local and nonlocal symmetries, and the Kac-Moody-Virasoro group invariant reductions are investigated.

Usually, starting from the trivial vacuum solution ($u=0$), the B\"acklund transformation will lead to one soliton solution. However, for the cKP3-4 equation \eqref{kp34}, the trivial vacuum solution and B\"acklund transformations will lead to abundant solutions including rational solutions, arbitrary D'Alembert type waves, solitons with a fixed form ($ \mbox{\rm sech}^2$ form) and
arbitrary velocity, and solitons  and soliton molecules with fixed velocity but arbitrary shapes (special D'Alembert waves).

There are two important (1+1)-dimensional symmetry reductions of the cKP3-4 equation \eqref{kp34}. The first type of reduction equation is Lax integrable with fourth order spectral problem. The second reduction is just the KdV equation. The more about the cKP3-4 equation \eqref{kp34} and its special reduction \eqref{RED1} will be reported in our future studies.

\section*{Acknowledgements\vspace{-0.2em}}
The work was sponsored by the National Natural Science Foundations of China (Nos.11975131,11435005) and K. C. Wong Magna Fund in Ningbo University.

\end{CJK*}
\end{document}